\documentclass[a4paper,11pt]{article}
\usepackage{pos}
\usepackage{subfloat}
\title{A search for spectral hardening in HAWC sources above 56 TeV}

\author*[a]{Kelly Malone}

\affiliation[a]{Los Alamos National Laboratory,\\
  Los Alamos, NM, USA}

\forColl{HAWC} 

\emailAdd{kmalone@lanl.gov}

\abstract{The High Altitude Water Cherenkov (HAWC) Observatory is a wide-field-of-view gamma-ray observatory that is optimized to detect gamma rays between ~300 GeV and several hundred TeV. The HAWC Collaboration recently released their third source catalog (3HWC), which contains 65 sources. One of these sources, the ultra-high-energy gamma-ray source 3HWC J1908+063, may exhibit a hardening of the spectral index at the highest energies (above 56 TeV). At least two populations of particles are needed to satisfactorily explain the highest energy emission. This second component could be leptonic or hadronic in origin. If it is hadronic in origin, it would imply the presence of protons with energies up to $\sim$1 PeV near the source. We have searched other 3HWC sources for the presence of this spectral hardening feature. If observed, this would imply that the sources could make good PeVatron candidates.}

\FullConference{37$^{\rm{th}}$ International Cosmic Ray Conference (ICRC 2021)\\
		July 12th -- 23rd, 2021\\
		Online -- Berlin, Germany}

\begin{document}
\maketitle

\section{Motivation}

3HWC J1908+063 is one of the brighest, highest-energy gamma-ray sources, with the High Altitude Water Cherenkov (HAWC) Observatory observing emission extending past 200 TeV~\cite{3hwc,lorentz,HECatalog}.

\begin{figure*}
\centering
\includegraphics[width=0.8\textwidth]{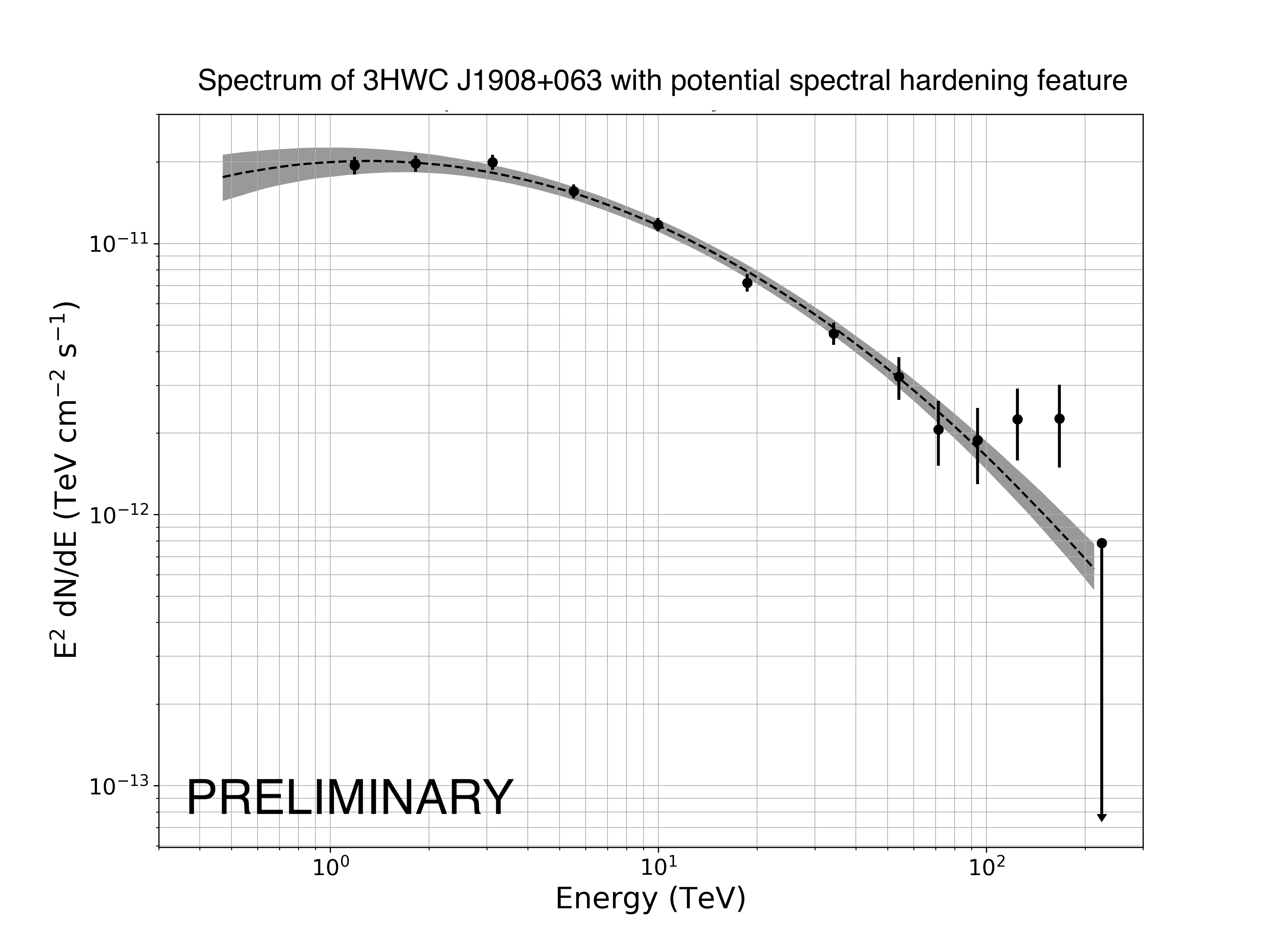}
\caption{The HAWC spectrum of 3HWC J1908+063, with the last few energy bins subdivided into smaller bins of equal width to more clearly see the spectral hardening feature.}.
\label{fig:j1908}
\end{figure*}

As discussed in ~\cite{j1908ICRC}, this source appears to show hints of spectral hardening above $\sim$ 75 TeV. This effect can be seen more clearly when the last three quarter-decade log-energy bins typically used by HAWC (as defined in ~\cite{Crab}) are subdivided into six smaller bins of equal width in log-energy space.  In the four significantly detected (TS $>$ 4.0) bins, the flux points appear roughly flat in E$^{2}$/dNdE space, deviating from the best-fit log-parabola spectrum. 

This effect is not presently significant. The last two flux points are a 1.5$\sigma$ and a 1.8$\sigma$ deviation from the best-fit line, respectively. Adding those two values in quadrature, the total significance is ~$\sim$2.3$\sigma$.  However, if shown to be significant with more data, this feature is potentially interesting as it may indicate that there are multiple populations of particles contributing to the TeV gamma ray emission. This shape is difficult to fit with single-population models. A hard spectrum at the highest energies could be indicative of hadronic emission. Searches for spectral hardening around 100 TeV could aid in identifying PeVatron candidates. 

HAWC is an extensive air shower array located at an altitude of 4100 meters in Puebla, Mexico. Its energy range extends past 100 TeV, and it has a wide field-of-view ($\sim$2 sr) that makes it a good instrument for performing surveys. Here, we search through HAWC's third catalog of sources (\cite{3hwc}, hereafter referred to as the ``3HWC catalog") to see if this spectral hardening feature is widespread among TeV gamma ray sources. 

\section{Method}

We downselect 65 sources reported in the 3HWC catalog to choose intriguing candidates for spectral hardening.  Sources must have a high enough significance that flux points can be obtained across the entire energy range. We impose a TS value of 50. 

The sources should also have an energy range that extends past 56 TeV. Table 2 of reference ~\cite{3hwc} contains the energy interval that is expected to contain 75$\%$ of a source's significance. This is not a perfect criterion; for example it is well-known that the Crab Nebula emits above 100 TeV~\cite{Crab} but it is removed by this cut because most of its significance comes from lower energy bins. Therefore, we cross-reference the list with HAWC's catalog of sources emitting above 56 TeV~\cite{HECatalog} and add in any sources that are missing. In practice, this only adds the Crab Nebula back into the list, as the rest of the highest-energy sources are selected using the first cut. Table \ref{tab:sources} shows the 19 sources of interest. 

\begin{table*}
\renewcommand{\arraystretch}{1.15}
\begin{center}
 \begin{tabular}{ | c | c | c | c | c |}
 \hline
 Source name & RA ($^{\circ}$) & Dec ($^{\circ}$) & TS & Energy range (TeV) \\
  \hline
  3HWC J0534+220 (Crab Nebula) & 83.63 & 22.01 & 35736.5 & 1.6 - 37.4 \\
  3HWC J0634+180 (Geminga region) & 98.75 & 18.05 & 36.2 & 3.7 - 102.0 \\
  3HWC J1809-190 & 272.46 & -19.04 & 264.8 & 7.7 - 177.3 \\
  3HWC J1813-125 & 273.34 & -12.52 & 51.9 & 2.6 - 69.2 \\
  3HWC J1813-174 & 273.43 & -17.47 & 416.0 & 7.7 - 174.7 \\
  3HWC J1819-150 & 274.79 & -15.09 & 93.8 & 2.2 -62.5 \\
  3HWC J1825-134 & 276.46 & -13.4 & 2212.5 & 9.2 - 183.4 \\
  3HWC J1831-095 & 277.87 & -9.59 & 237.7 & 4.2 - 106.7 \\
  3HWC J1837-066 & 279.40 & -6.62 & 1542.7 & 2.2 - 57.3 \\
  3HWC J1843-034 & 280.99 & -3.47 & 876.6 & 6.2 - 142.6 \\
  3HWC J1849+001 & 282.35 & 0.15 & 427.5 & 9.9 - 195.3 \\
  3HWC J1908+063 & 287.05 & 6.39 & 1320.9 & 8.9 - 182.7 \\
  3HWC J1922+140 & 290.70 & 14.09 & 176.6 & 2.1 - 60.0 \\
  3HWC J1928+178 & 292.10 & 17.82 & 216.7 & 5.9 - 140.5 \\
  3HWC J1951+293 & 297.99 & 29.40 & 68.7 & 4.0 - 108.6 \\
  3HWC J2006+340 & 301.73 & 34.00 & 67.4 & 3.3 - 83.5 \\
  3HWC J2019+367 & 304.94 & 36.80 & 1227.5 & 11.7 - 211.7 \\
  3HWC 2031+415 (Cygnus Cocoon region) & 307.93 & 41.51 & 556.9 & 6.3 - 147.8 \\
  3HWC J2227+610 & 336.96 & 61.05 & 52.5 & 14.3 - 292.7 \\
  \hline
\end{tabular}
\caption{The sources selected for the analysis. Adapted from ~\cite{3hwc}. TS refers to the test statistic from the likelihood fit. The energy range is the interval containing 75$\%$ of the source's significance.}\label{tab:sources}
\end{center}
\end{table*}

The procedure to search for spectral hardening is as follows: first, each region is fit using HAWC's nominal energy bins (quarter-decade widths in log-energy space) to determine the best spectral shape and morphology. The ``ground parameter" energy estimator is used~\cite{Crab}.  The free parameters in the spectral and morphological models are simultaneously fit via a likelihood fit. The HAL (HAWC Accelerated Likelihood)\footnote{\url{https://github.com/threeML/hawc_hal}} plugin to the 3ML (Multi-mission Maximum Likelihood) framework~\cite{threeml} is used. The definitions of the morphological and spectral shapes are contained in the \textit{astromodels}\footnote{\url{https://github.com/threeML/astromodels}} software package.  Three different spectral shapes are considered: a power-law, a power-law with an exponential cutoff, and a log-parabola. The Bayesian information criterion for each likelihood fit are then compared to determine which spectral shape provides the best fit to the data. Spectral points are then  obtained using the procedure detailed in ~\cite{Crab}. 

Some of the sources from Table \ref{tab:sources} have been the subject of dedicated follow-up papers by the HAWC Collaboration. For those sources, we deviate from the procedure above. If the source has previously been studied more in depth, the spectrum and morphology from the dedicated analysis is used. For example, 3HWC J2031+415 has been resolved into multiple sources. As discussed in ~\cite{Cocoon}, this region actually consists of a large extended source (the Cygnus cocoon) along with a high-energy pulsar. We simply use the reported best-fit spectral shapes and morphology from ~\cite{Cocoon}.

When performing the fits, 3HWC sources within 2.5 degrees of the source of interest are included in the model. This reduces contamination from nearby sources.

After the best spectral shape and morphology are determined, each source is fit again. The last three quarter-decade log-energy bins, corresponding to energies above 56 TeV, are subdivided into six smaller bins of equal length. The boundaries of each bin are reported in Table \ref{tab:energies}.  Sub-dividing the highest energy bins allows for a better energy resolution and makes it easier to see if spectral hardening is present. 

\begin{table*}
\renewcommand{\arraystretch}{1.15}
\begin{center}
 \begin{tabular}{ | c | c | c | }
 \hline
 Bin name & $E_l$ (TeV) & $E_h$ (TeV) \\
  \hline
  a & 0.316 & 0.562 \\
  b & 0.562 & 1.00 \\
  c & 1.00 & 1.78 \\
  d & 1.78 & 3.16 \\
  e & 3.16 & 5.62 \\
  f & 5.62 & 10.0 \\
  g & 10.0 & 17.8 \\
  h & 17.8 & 31.6 \\
  i & 31.6 & 56.2 \\
  j$_1$ & 56.2 & 75.0 \\
  j$_2$ & 75.0 & 100 \\
  k$_1$ & 100 & 133 \\
  k$_2$ & 133 & 177 \\
  l$_1$ & 177 & 234 \\
  l$_2$ & 234 & 316 \\
  \hline
\end{tabular}
\caption{The energy bin boundaries for the flux points reported in Section 3. $E_l$ and $E_h$ are the low and high values, respectively. They are different from the typical HAWC energy bins; bins above 56 TeV are narrower to allow for better energy resolution when searching for spectral hardening. This analysis is restricted to reconstructed energies above 1 TeV, so bins $a$ and $b$ are not used. }\label{tab:energies}
\end{center}
\end{table*}

We then search for hints of spectral hardening by calculating how much the flux points deviate from the best-fit spectrum. 

\section{Selected results}

\begin{figure*}
\includegraphics[width=0.37\textwidth]{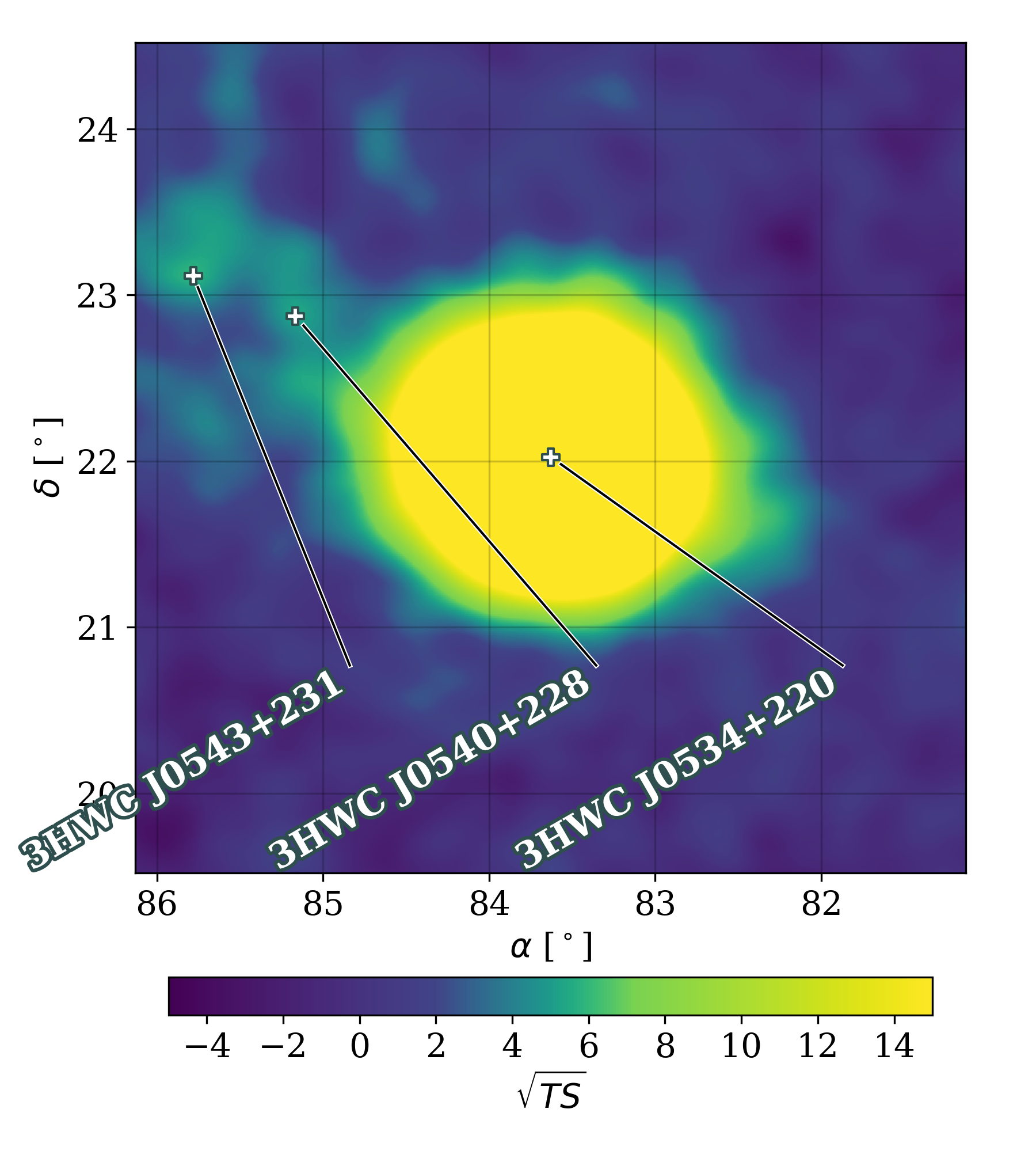}
\includegraphics[width=0.63\textwidth]{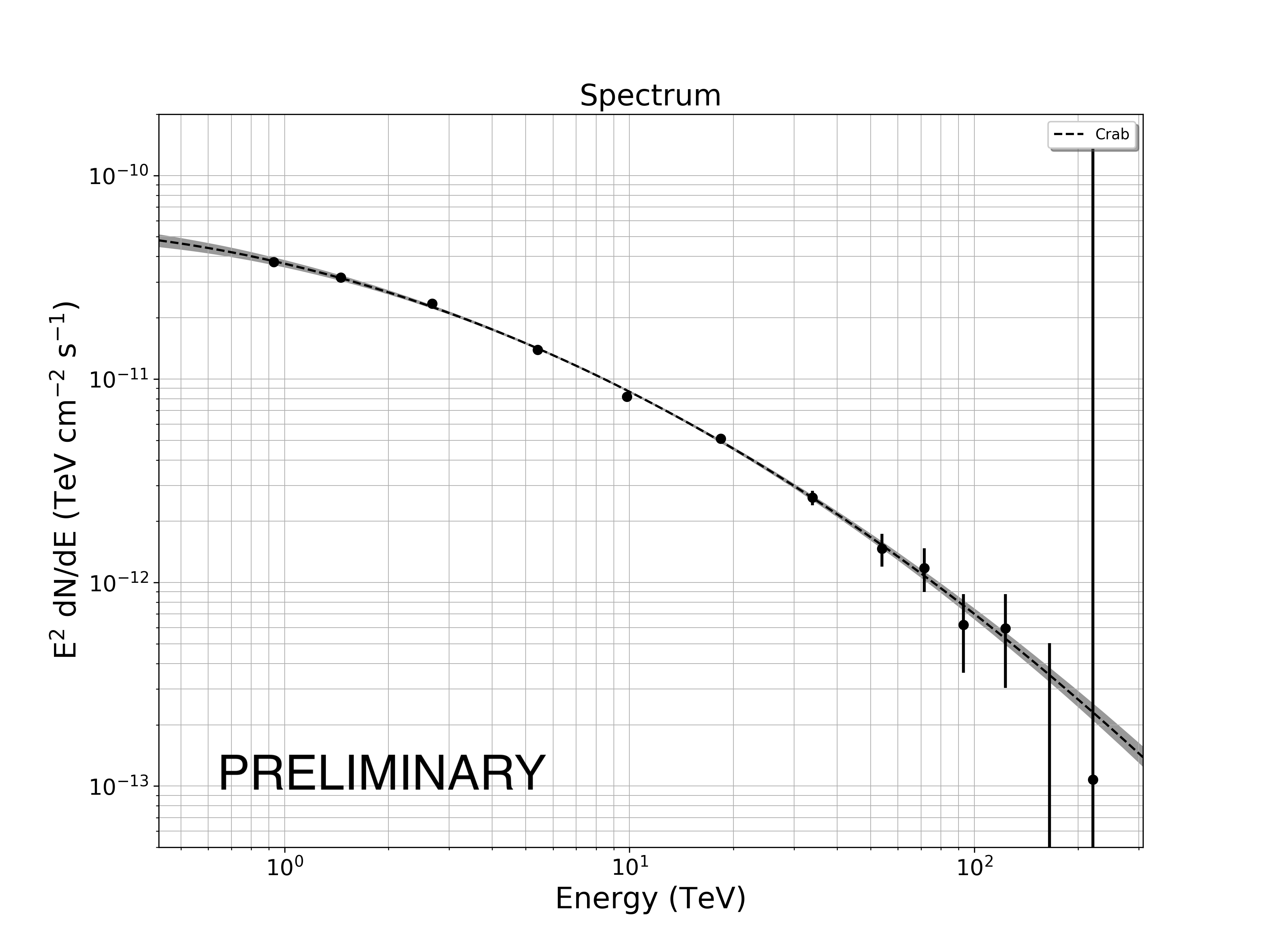}
\caption{Left: HAWC significance map of the region around 3HWC J0534+220, the Crab Nebula. Right: The spectrum of the Crab Nebula. No spectral hardening is observed.} 
\label{fig:crab}
\end{figure*}

\begin{figure*}
\includegraphics[width=0.37\textwidth]{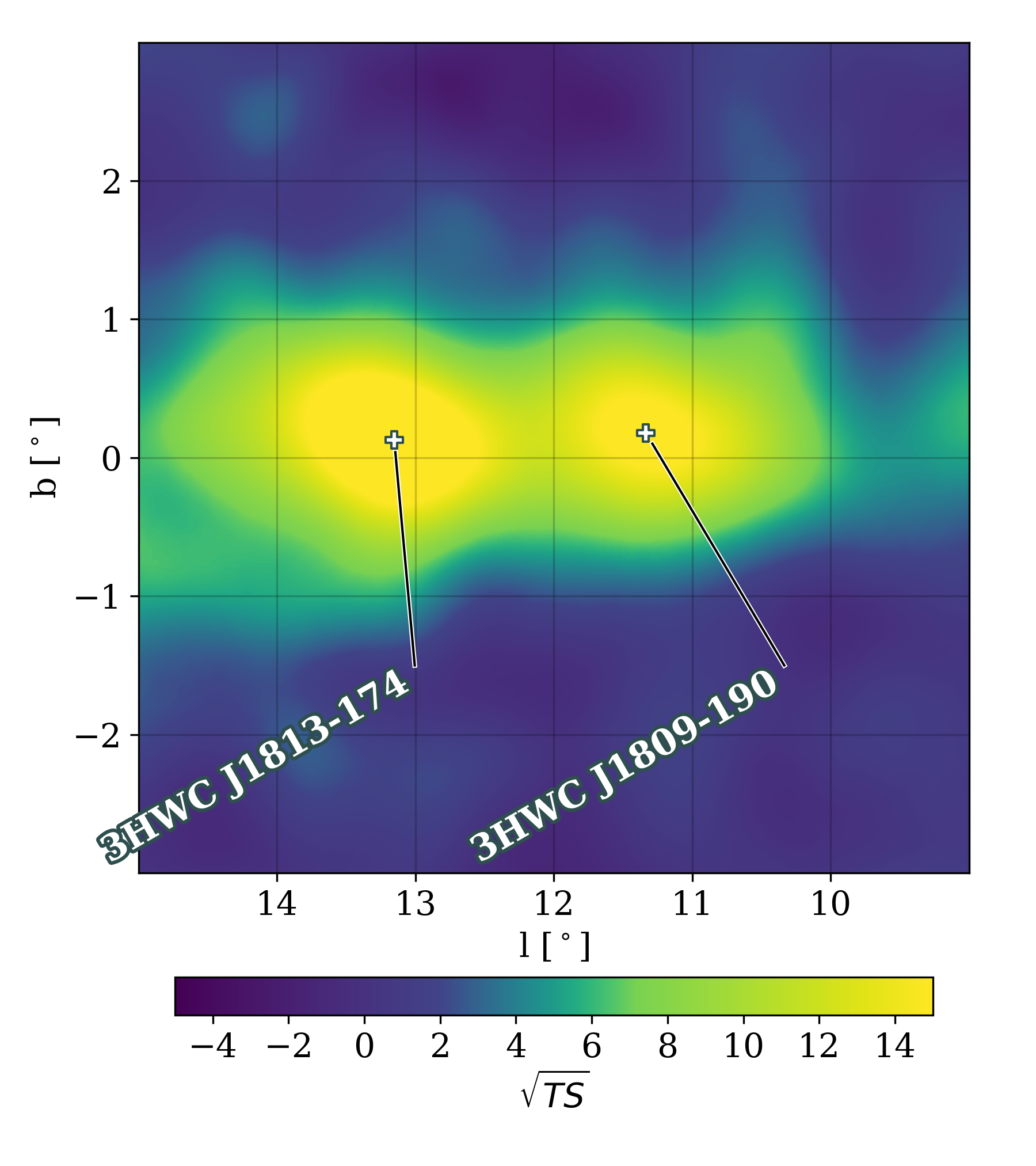}
\includegraphics[width=0.63\textwidth]{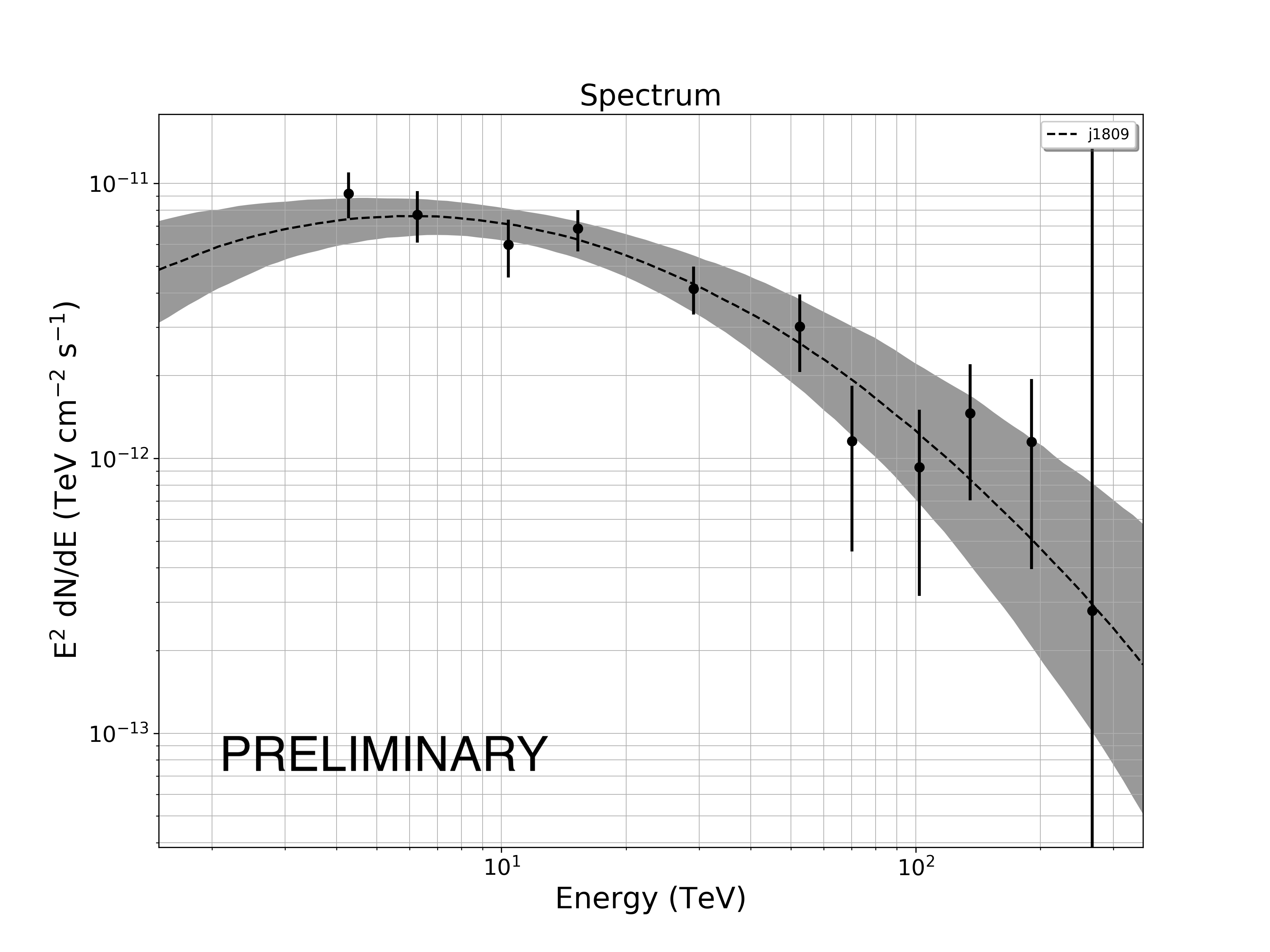}
\caption{Left: HAWC significance map of the region around 3HWC J1809-190. Right: The spectrum of 3HWC J1809-190} 
\label{fig:j1809}
\end{figure*}

\begin{figure*}
\includegraphics[width=0.37\textwidth]{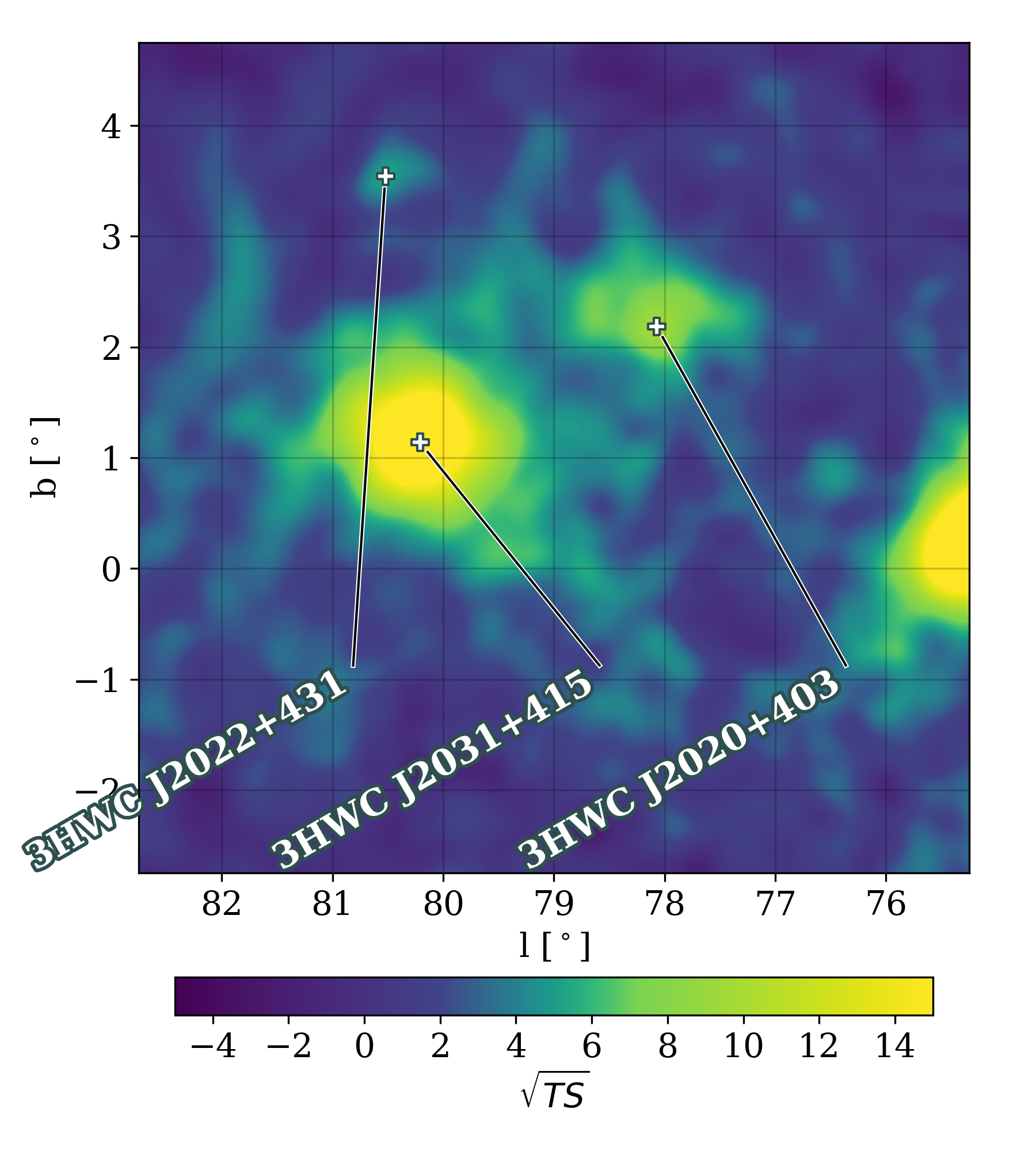}
\includegraphics[width=0.63\textwidth]{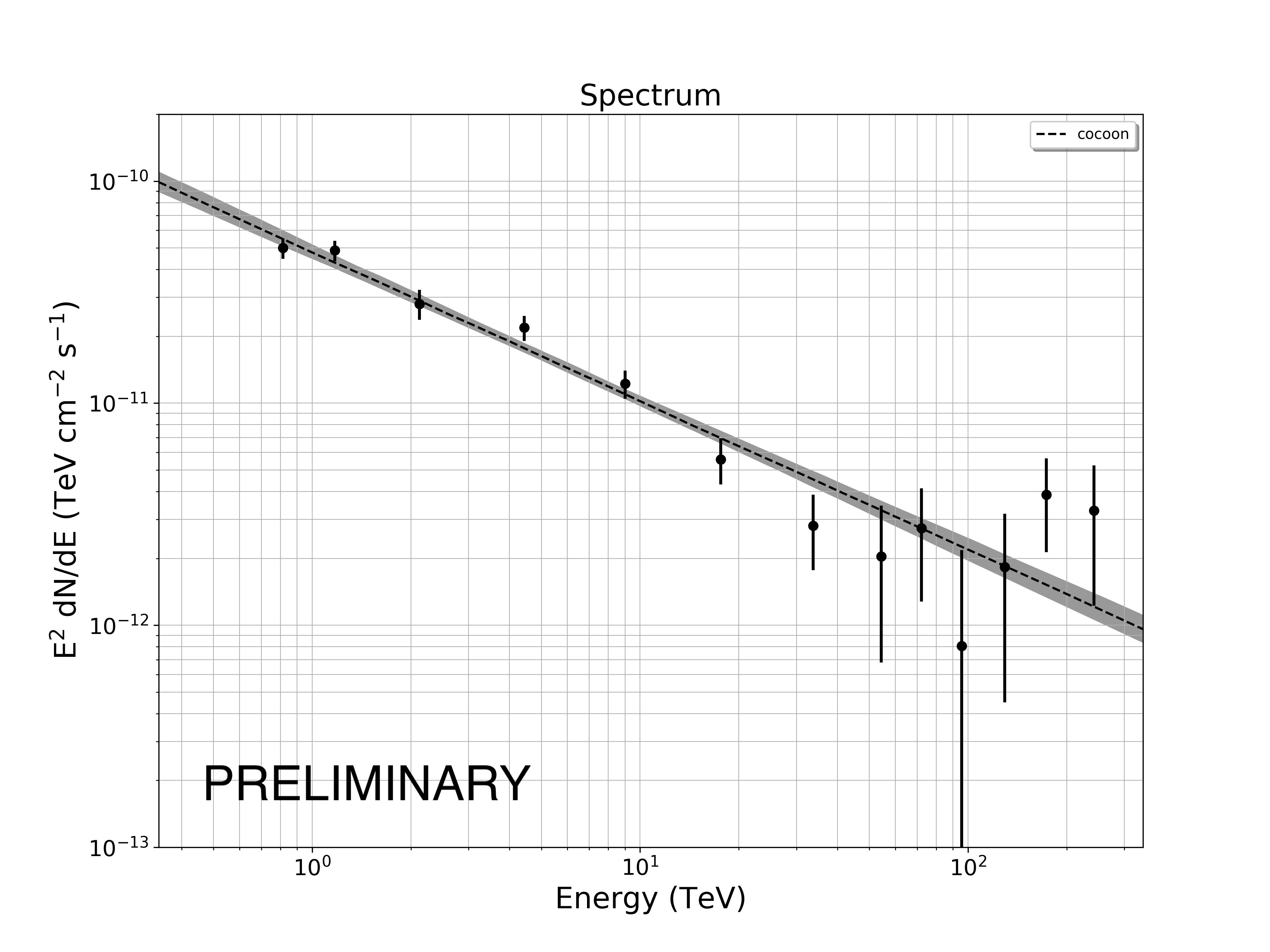}
\caption{Left: HAWC significance map of the region around 3HWC J2031+415. Right: The spectrum of the Cygnus Cocoon, which is a large, angularly extended superbubble surrounding a region of massive star formation and is contained in the region. } 
\label{fig:cocoon}
\end{figure*}

Selected results are shown in Figures \ref{fig:crab} through \ref{fig:cocoon}.

Figure \ref{fig:crab} shows the significance map and spectrum of the Crab Nebula. This source is commonly used as a standard candle in gamma-ray astrophysics. No evidence of spectral hardening is observed. The last two flux points have very low TS values and the uncertainties are vey large. The rest of the flux points lie right along the best-fit log-parabola line. This shows that the feature observed in 3HWC J1908+063 is likely not an instrumental effect related to mis-modeling the effective area of the HAWC detector at the highest energies. 

The other two sources show some of the more promising candidates for spectral hardening. 

Figure \ref{fig:j1809} shows the significance map of the 3HWC J1809-190/3HWC J1813-174 region. Due to the proximity of these sources to each other, they are fit simultaneously. The figure of 3HWC J1809-190 is also shown. The source is not significantly detected in the last bin; the TS value is 0.30. This explains why the uncertainty on that flux point is so high.  The two bins before that, bins m and n, have TS values of 9.6 and 5.6, respectively. Each of these flux points deviate from the best-fit log-parabola line by approximately 1$\sigma$. Adding these two points in quadrature, the total amount of the deviation is approximately 1.4$\sigma$.

Figure \ref{fig:cocoon} shows the significance map around the 3HWC J2031+415 region, along with the flux points for the Cygnus Cocoon source. By eye, it looks like the spectrum may be flattening out, which a slight deviation from the best-fit powerlaw spectrum. The last two flux points have TS values of 5.7 and 3.4, respectively, and deviate from the best-fit spectrum by 1.4$\sigma$ and 1.1$\sigma$, respectively. This gives a total deviation of 1.8$\sigma$.

\section{Conclusions} 

As discussed in ~\cite{j1908ICRC}, the spectrum of 3HWC J1908+063 may exhibit spectral hardening at the highest energies. The deviation from the best-fit spectrum is approximately 2.3$\sigma$. This is potentially interesting as it could aid in determining the origins of the emission.

Here, we search through sources in the 3HWC catalog to see if indiciations of spectral hardening are widespread. Many of these sources are much dimmer than 3HWC J1908+063. No significant spectral hardening is observed. The most significant sources exhibit deviations from their best-fit spectra of less than 2$\sigma$.

Upgraded HAWC reconstruction algorithms, especially those that provide for better gamma/hadron separation and/or energy resolution, could aid in determining whether spectral hardening is present. A re-analysis with data including HAWC's recently completed outrigger array could be particularly informative. Systematic uncertainties, both related to mis-modeling of the detector and due to potential source confusion, need to be investigated in-depth. This work is in progress.

\acknowledgments

We acknowledge the support from: the US National Science Foundation (NSF); the US Department of Energy Office of High-Energy Physics; the Laboratory Directed Research and Development (LDRD) program of Los Alamos National Laboratory; Consejo Nacional de Ciencia y Tecnolog\'ia (CONACyT), M\'exico, grants 271051, 232656, 260378, 179588, 254964, 258865, 243290, 132197, A1-S-46288, A1-S-22784, c\'atedras 873, 1563, 341, 323, Red HAWC, M\'exico; DGAPA-UNAM grants IG101320, IN111716-3, IN111419, IA102019, IN110621, IN110521; VIEP-BUAP; PIFI 2012, 2013, PROFOCIE 2014, 2015; the University of Wisconsin Alumni Research Foundation; the Institute of Geophysics, Planetary Physics, and Signatures at Los Alamos National Laboratory; Polish Science Centre grant, DEC-2017/27/B/ST9/02272; Coordinaci\'on de la Investigaci\'on Cient\'ifica de la Universidad Michoacana; Royal Society - Newton Advanced Fellowship 180385; Generalitat Valenciana, grant CIDEGENT/2018/034; Chulalongkorn University’s CUniverse (CUAASC) grant; Coordinaci\'on General Acad\'emica e Innovaci\'on (CGAI-UdeG), PRODEP-SEP UDG-CA-499; Institute of Cosmic Ray Research (ICRR), University of Tokyo, H.F. acknowledges support by NASA under award number 80GSFC21M0002. We also acknowledge the significant contributions over many years of Stefan Westerhoff, Gaurang Yodh and Arnulfo Zepeda Dominguez, all deceased members of the HAWC collaboration. Thanks to Scott Delay, Luciano D\'iaz and Eduardo Murrieta for technical support.


\bibliographystyle{JHEP}
\bibliography{bib}


\clearpage
\section*{Full Authors List: \Coll\ Collaboration}

\scriptsize
\noindent
A.U. Abeysekara$^{48}$,
A. Albert$^{21}$,
R. Alfaro$^{14}$,
C. Alvarez$^{41}$,
J.D. Álvarez$^{40}$,
J.R. Angeles Camacho$^{14}$,
J.C. Arteaga-Velázquez$^{40}$,
K. P. Arunbabu$^{17}$,
D. Avila Rojas$^{14}$,
H.A. Ayala Solares$^{28}$
% ...

\noindent
% ...
$^{14}$Instituto de Física, Universidad Nacional Autónoma de México, Ciudad de México, México,
% ...
$^{17}$Instituto de Geofísica, Universidad Nacional Autónoma de México, Ciudad de México, México,
% ...
$^{21}$Physics Division, Los Alamos National Laboratory, Los Alamos, NM, USA,
% ...
$^{28}$Department of Physics, Pennsylvania State University, University Park, PA, USA,
% ...
$^{40}$Instituto de Física y Matemáticas, Universidad Michoacana de San Nicolás de Hidalgo, Morelia, Michoacán, México,
$^{41}$FCFM-MCTP, Universidad Autónoma de Chiapas, Tuxtla Gutiérrez, Chiapas, México,
% ...
$^{48}$Department of Physics and Astronomy, University of Utah, Salt Lake City, UT, USA,
% ...

\end{document}